\documentclass[sigconf,nonacm]{acmart}

\usepackage{multirow}
\usepackage{float}
\usepackage[utf8]{inputenc}
\usepackage{wrapfig}
\usepackage{graphicx}
\usepackage{xcolor}
\usepackage{xspace}
\usepackage[htt]{hyphenat}

\usepackage{colortbl}

\AtBeginDocument{%
  }

\setcopyright{acmlicensed}
\copyrightyear{2025}
\acmYear{2025}
\acmDOI{XXXXXXX.XXXXXXX}

\acmConference[CHI '25]{CHI conference on Human Factors in Computing Systems}{April 26--May 1, 2025}{Yokohama, Japan}

\newcommand{\code}[1]{\texttt{#1}}

\newcommand{\ethics}{Paris School of Economics ethical committee (2024-007)}
\usepackage{float}
\begin{document}

\title{The Moral Turing Test: Evaluating Human-LLM Alignment in Moral Decision-Making }

\author{Basile Garcia}
\orcid{0000-0002-3227-3471}
\email{basile.garcia@unige.ch}
\affiliation{%
  \institution{University of Geneva}
  \city{Geneva}
  \country{Switzerland}
}

\author{Crystal Qian}
\orcid{0000-0001-7716-7245}
\email{cjqian@google.com}
\affiliation{%
  \institution{Google DeepMind}
  \city{New York City}
  \state{NY}
  \country{USA}
}

\author{Stefano Palminteri}
\orcid{0000-0001-5768-6646}
\email{stefano.palminteri@ens.fr}
\affiliation{%
  \institution{École Normale Supérieure}
  \city{Paris}
  \country{France}
}

\begin{abstract}
As large language models (LLMs) become increasingly integrated into society, their alignment with human morals is crucial. To better understand this alignment, we created a large corpus of human- and LLM-generated responses to various moral scenarios. We found a misalignment between human and LLM moral assessments; although both LLMs and humans tended to reject morally complex utilitarian dilemmas, LLMs were more sensitive to personal framing. We then conducted a quantitative user study involving 230 participants (N=230), who evaluated these responses by determining whether they were AI-generated and assessed their agreement with the responses. Human evaluators preferred LLMs' assessments in moral scenarios, though a systematic anti-AI bias was observed: participants were less likely to agree with judgments they believed to be machine-generated. Statistical and NLP-based analyses revealed subtle linguistic differences in responses, influencing detection and agreement. Overall, our findings highlight the complexities of human-AI perception in morally charged decision-making.
\end{abstract}

\received{12 September 2024}

\maketitle
\section{Introduction}
Large language models (LLMs) are becoming widely used in applications ranging from conversational agents to decision-making systems capable of making consequential decisions, such as providing medical advice \cite{haupt_ai-generated_2023}, legal advice \cite{cheong_i_2024}, and mental well-being support \cite{marrapese_novel_2024}. As humans increasingly interact with LLMs, understanding our ability to detect and align with LLMs' judgments becomes crucial, particularly given the risk of misuse, such as the dissemination of disinformation by LLM-powered bots \cite{doshi_sleeper_2024, sun_exploring_2024}.

While prior work has explored AI detection and alignment, the relationship between identification and agreement remains empirically under-investigated, especially in moral decision-making processes. Specifically, it remains unclear whether a participant’s belief about the source of content affects their agreement. Our research directly addresses this gap, exploring humans' capacity to detect the source of moral judgments (either human or LLM), their agreement with these judgments, and critically, the relation between these two behavioral outcomes. Additionally, we explore the linguistic factors influencing identification and agreement \cite{mitrovic_chatgpt_2023, scherrer_evaluating_2023}.

To investigate human-AI alignment in moral decision-making, we conducted a series of quantitative experiments involving 230 participants (N=230). First, we collected a corpus of moral judgments by presenting 60 diverse ethical scenarios to human participants and LLM models in the GPT-3.5 family. We then presented these judgments to a new group of participants, who were tasked with identifying the source (human or AI), expressing their agreement or disagreement with the judgment itself, as well as agreement with the accompanying justification. To control for detection bias, we created and evaluated additional corpora with ``humanized'' LLM responses. Here are the key highlights from our analysis:

\begin{itemize}
    \item \textbf{LLMs exhibit a different moral code from humans, and from each other}: We found that LLMs are highly sensitive to personal vs. impersonal framing; GPT-3.5 \code{davinci-text-003}, in particular, was much more likely to agree with actions taken in impersonal moral scenarios (where they do not bear personal responsibility) than in personal moral reframings (where they bear personal responsibility). Furthermore, we found that the framing effect was greatly exacerbated between GPT-3.5 \code{davinci-text-002} and GPT-3.5 \code{davinci-text-003}, suggesting that moral judgments may be model-dependent. 

    \item \textbf{Participants prefer AI justifications over human justifications in morally-complex scenarios}: Although participants preferred human justifications when the stakes were low (e.g., in non-moral scenarios), they significantly preferred LLM-generated justifications in personal moral scenarios (such as when explaining how they would handle the trolley problem), where LLMs exhibited much stronger utilitarian preferences than humans. Participants' preference for AI in these scenarios may stem from a preference for deliberative reasoning in high-stakes settings.
    
    \item \textbf{However, participants exhibit a strong anti-AI bias}: Even though participants favored the justifications produced by LLMs, they reported disagreement if they suspected that the output was LLM-generated. Across all types of scenarios, participants exhibited a notable anti-AI bias. This result is robust to our efforts to conceal the identity of the LLM through ``humanizing'' linguistic features, such as introducing typos.
    
    \item \textbf{Subtle contextual and linguistic cues can reveal AI authorship}: Participants were able to detect the source of generated justifications with moderate accuracy. The detection rate was higher in moral scenarios (such as the trolley problem) than in non-moral scenarios. Slight linguistic differences, such as an increased use of first-person pronouns in human explanations and more pedantic, analytical LLM-generated explanations, provided some signal.
\end{itemize}

\section{Background}
\subsection{Human moral psychology}
Moral psychology investigates how people make ethical decisions and evaluate others' actions. Research indicates that moral judgments are often driven by immediate emotions rather than deliberate reasoning \cite{lapsley_moral_2018}. For example, in the trolley problem, individuals are asked if they would sacrifice one person to save five. Responses vary depending on whether the scenario is framed \textit{personally} (where one must actively push the person onto the tracks) or \textit{impersonally}. This suggests that moral judgments are frequently inconsistent, influenced by context and cognitive biases \cite{greene_fmri_2001, cushman_role_2006}.

Dual process theories offer a popular explanation for these inconsistencies \cite{neys_dual_2006,kahneman_thinking_2011}. These theories propose that moral judgment relies on two competing cognitive systems: one that is fast, intuitive, and emotion-driven (“hot”), and another that is slow, deliberative, and rational (“cold”) \cite{greene_neural_2004, koenigs_damage_2007}. The deliberative system follows utilitarian principles, focusing solely on the outcomes of decisions, while the intuitive system is swayed by contextual factors unrelated to the final outcome, leading to automatic, emotional responses. Consequently, moral preferences can be inconsistent, as different framing of similar outcomes trigger varying responses \cite{kahneman_choices_1984, sinnott-armstrong_framing_2008}.

\subsection{AI moral psychology}

Moral scenarios can also be used to study ethics and alignment within AI systems \cite{kazim_high-level_2021, hendrycks_aligning_2021}. As LLMs increase their capacity for conversational decision-making, the practice of recycling tools from cognitive psychology to study LLMs' competencies in terms of decision-making and reasoning has emerged. Several recent studies took the challenge to recycle tools from cognitive psychology to the study of LLMs’ competences in terms of decision-making and reasoning \cite{binz_using_2023, hagendorff_human-like_2023, yax_studying_2024}. 

Scherrer et. al. finds that LLMs generally align with human moral values, but in ambiguous cases, their responses can vary based on question phrasing, with closed-source models demonstrating more consistent preferences \cite{scherrer_evaluating_2023}. This variation may arise from differences in pre-training data and fine-tuning processes \cite{park_correct_2023, zhu_language_2024}. Thus, linguistic features play a significant role in assessing the moral quality of judgments from humans or AI.

\subsection{Factors influencing AI detection}

Determining whether a decision is made by a human or a machine is crucial: it enhances safety by revealing our susceptibility to manipulation, acts as an epistemological Turing test~\cite{jones_does_2024} for assessing AI conversational abilities, and guides the development of LLMs towards human-preferred outputs~\cite{chiang_chatbot_2024}. Concerningly, recent studies indicate that humans often struggle to reliably distinguish AI-generated texts  from human texts, across diverse contexts such as poetry~\cite{clark_all_2021, kobis_artificial_2021-1} and media misinformation \cite{kreps_all_2020}. Additionally, strategies can be employed to ``humanize'' AI-generated content to increase the difficulty of detection. ``Humanized'' LLMs have sometimes been judged as more human-like than actual human-generated responses~\cite{jakesch_human_2023}, and LLMs can be perceived as more empathetic than human responses when prompted appropriately~\cite{welivita_are_2024}.

\subsection{Factors influencing AI alignment}

Before LLMs, research into applied fields such as autonomous vehicles highlighted the need for alignment in human and machine moral decision making~\cite{awad_moral_2018}. Prior research has shown a human tendency to favor human-generated decisions over machine-generated ones, a phenomenon known as algorithm aversion \cite{burton_systematic_2020}. However, this phenomenon is context-dependent \cite{castelo_task-dependent_2019}. For instance, humans tend to prefer human judgement over AI judgement in the context of medical decision-making~\cite{cadario_understanding_2021}, but prefer AI judgement in numerical tasks~\cite{logg_algorithm_2019}.

Agreement with LLM-generated text hinges on factors like task nature, perceived authorship \cite{de_freitas_psychological_2023}, prior AI interactions, and even cultural context \cite{kapania_because_2022}. While ChatGPT's responses in social scenarios have been rated as more balanced and empathetic than human advice \cite{howe_chatgpts_2023}, people still show a strong preference for human advice on moral issues \cite{proksch_impact_2024}. AI authors are perceived as less competent, though humans still value their advice \cite{bohm_people_2023-1}. Moreover, in persuasive content creation (e.g. advertisements), AI efforts are often rated higher than human efforts. Revealing the source of content production lessens the quality gap between human- and AI-generated content, without affecting the assessed quality of AI-created content \cite{zhang_human_2023}. This suggests that human favoritism, rather than AI aversion, drives the perceived quality and perceived value \cite{morewedge_preference_2022}.

To summarize, two competing hypotheses can help explain perceptions of LLMs in moral decision-making. A pro-AI bias may occur when machines are seen as authoritative sources of knowledge. In contrast, an anti-AI bias might emerge from societal or psychological prejudices against machines, often stemming from the belief that machines lack agency and the capacity for compassionate or morally sound decisions \cite{bigman_people_2018}.

\section{Methods}

\begin{figure*}
    \centering
    \includegraphics[width=.7\textwidth]{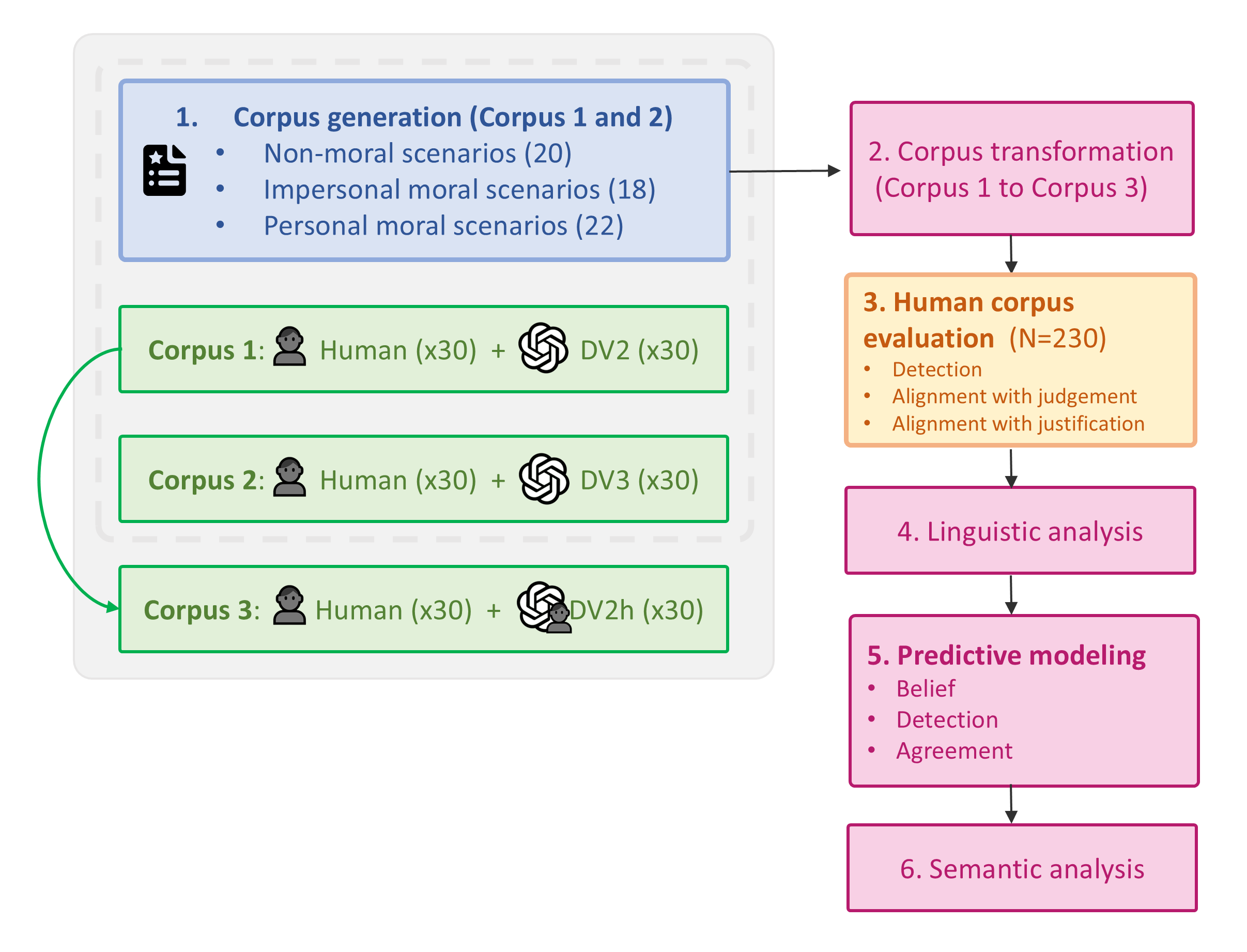}
    \caption{A diagram showing the experimental design. The blue and orange boxes (Steps 1 and 3) correspond to respectively Corpus generation and Detection and Agreement experiments (see Figure 2). The purple boxes (Steps 2, 4, 5, and 6) denote quantitative, statistical, and computational methods. The output of the experiment is an analysis on three corpora of human- and LLM- generated responses to various scenarios.}
    \label{fig:0}
\end{figure*}

Our experimental design is summarized follows (and shown in Figure \ref{fig:0}).\footnote{The research was conducted in accordance with the principles and guidelines for experiments involving human participants as outlined in the Declaration of Helsinki (1964, revised in 2013). The study received approval from \ethics. Informed consent was obtained from all participants prior to their involvement in each experiment.}

\begin{enumerate}
    \item \textbf{Corpus generation} First, we create two corpora of responses to scenarios of various types: non-moral, impersonal moral, or personal moral. For each scenario in corpus 1, 30 human participants and 30 API calls of GPT 3.5 \code{davinci-text-002} (\textbf{\code{dv2}}) provide a \textit{response}, which includes a \textit{judgement} (yes/no) and a \textit{justification} (free text). Corpus 2 uses GPT 3.5 \code{davinci-text-003} (\textbf{\code{dv3}}) instead of \code{dv2}.
    \item \textbf{Corpus transformation} Because human- and LLM- generated justifications may have liguistic differences (such as typos, or response length), we create corpus 3, which ``humanizes'' the \code{dv2} responses from corpus 1 by adding typos or shortening the text.
    \item \textbf{Corpus evaluation} Next, we have human raters evaluate each response from the 3 corpora. For each evaluation, they answer 1) whether they think the text was human- or LLM- generated, 2) whether they agree with the \textit{judgement}, and 3) whether they agree with the \textit{justification}.
    \item \textbf{Linguistic analysis} These remaining steps use statistical and computational analysis to figure out which specific signals are being communicated in the justification text to affect detection and alignment. In this step, we perform a linguistic analysis on potential linguistic differences between human- and LLM- generated text.
    \item \textbf{Predictive modeling} We use NLP techniques to evaluate whether the responses in the corpora have predictive signals on our outcomes of interest. We build models to predict 1) the true source of the text, 2) the human rater's belief of the source of the text, 3) whether the human rater's belief was correct, and 4) whether the human agrees with the justification. 
    \item \textbf{Semantic analysis} Finally, we use interpretability methods to identify \textit{which specific tokens} in the justification text have predictive power on the outcomes in described in 4.
\end{enumerate}

\subsection{Corpus generation}
In the preliminary item-generating stage of this study (Figure \ref{fig:methods-corpus}A; corpus generation experiments), we presented 60 moral scenarios from Greene et al. \cite{greene_neural_2004-1} to participants. There were three categories of these scenarios:
\begin{itemize}
    \item \textbf{Non-moral}: Scenarios that do not engage complex moral reasoning, e.g. ``Is it appropriate to wait for a promotion before buying a new computer, versus buying it immediately?''
    \item \textbf{Impersonal moral}: Moral scenarios involving indirect intervention, e.g. ``Is it appropriate to turn a runaway boxcar away from five people but towards one person?''
    \item \textbf{Personal moral}: Emotionally salient scenarios that are more engaging, e.g. ``Is it appropriate to push one person off a bridge to stop a runaway boxcar from hitting five people?'' These scenarios are similar to impersonal moral scenarios but involve a sense of personal agency.
\end{itemize}

Participants provided a yes/no decision and a free-text justification for each scenario. Each corpus contained 3600 responses, derived from 60 scenarios with 30 human responses and 30 LLM responses per scenario. The temperature of the LLM calls was set to 0.7, to maintain a form of inter-prompt variability similar to what is observed in humans.  After removing ambiguous answers, we created two corpora:

\begin{itemize}
    \item \textbf{Corpus 1}: 3542 responses from the \code{davinci-text-002} batch of experiments. Contains human and \code{dv2} responses.
    \item \textbf{Corpus 2}: 3420 responses from the \code{davinci-text-003} batch of experiments. Contains human and \code{dv3} responses.
\end{itemize}

\subsection{Corpus transformation}
\begin{figure*}
    \centering
    \includegraphics[width=\textwidth]{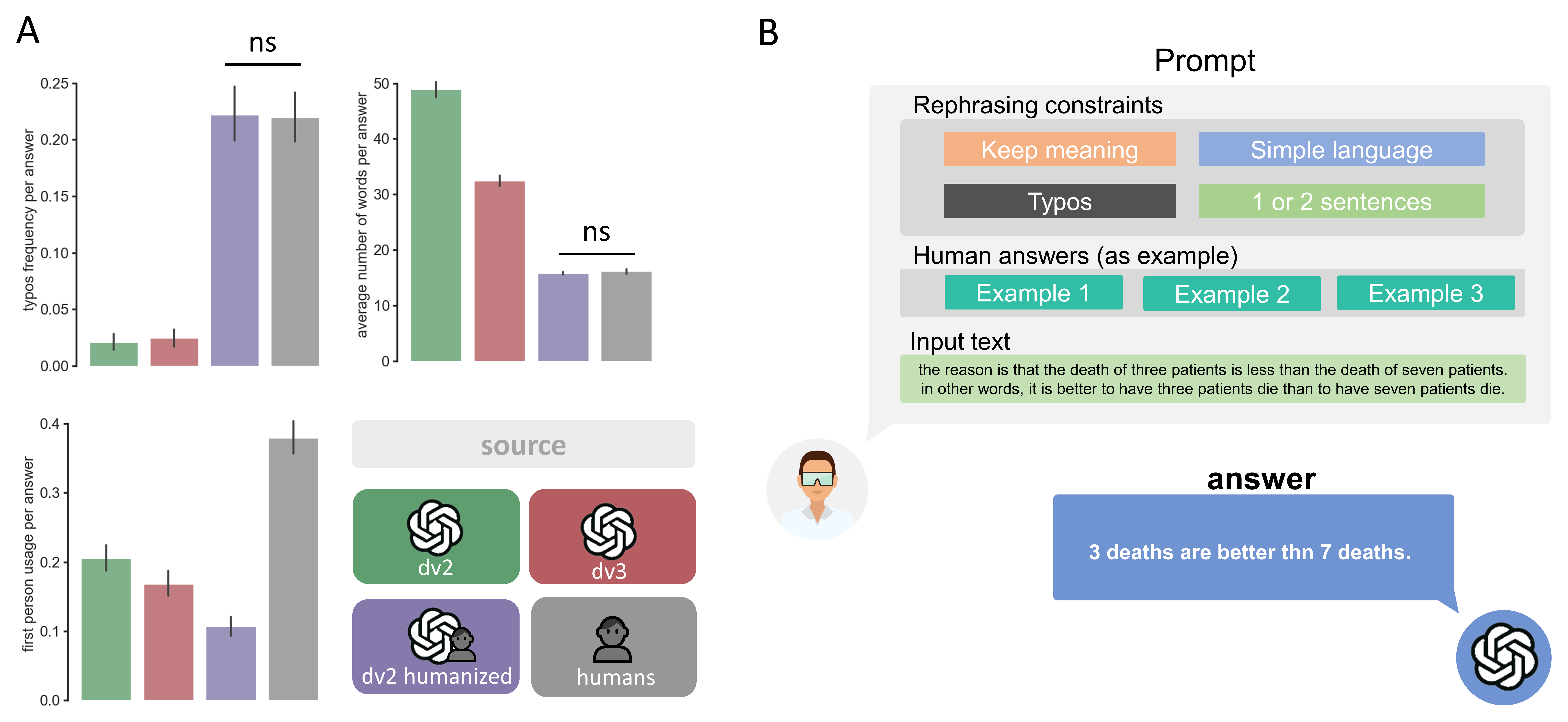}
    \caption{(A) Identified linguistic features which have been found to be different between human- and LLM-generated responses. (dv2: text-davinci-002; dv3: text-davinci-003, dv2h: humanized dv2). (B) Schematized prompting strategy to generate the humanized LLM response, by reducing size and including typos.}
    \label{fig:linguistic-diffs}
\end{figure*}

To mitigate any potential linguistic differences, we used GPT-3.5 to "humanize" the LLMs' responses in corpus 1 by shortening them, introducing a few typos, and providing human examples to imitate; this process is shown in Figure \ref{fig:linguistic-diffs}. Our resulting \code{dv2\_humanized} dataset showed no significant difference in length or number of typos from the human dataset, as evidenced by statistical t-tests. This resulting dataset of human and \code{dv2\_humanized} responses is \textbf{corpus 3}.

\subsection{Corpus evaluation}
\begin{figure*}
    \centering
    \includegraphics[width=\textwidth]{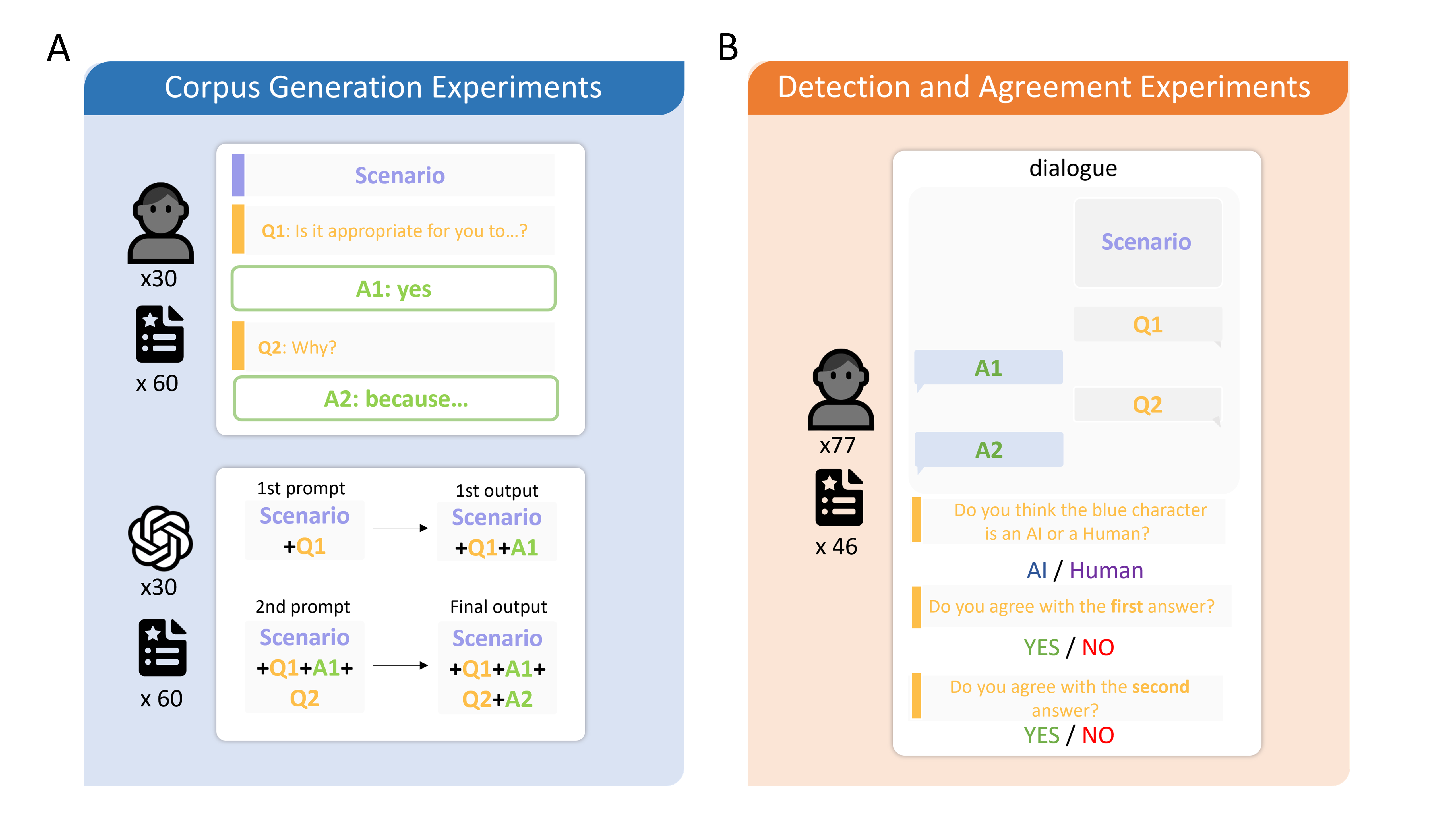}
    \caption{ (A) Schematic interface and method used in the experiment we used to generate corpus 1 (human and \code{dv2} responses) and corpus 2 (human and \code{dv3} responses). (B) Schematic interface used in the \textit{3. corpus evaluation} step.}
    \label{fig:methods-corpus}
\end{figure*}

In the main stage of the study (Figure \ref{fig:methods-corpus}B), we presented the responses generated in the first stage to new human participants (N=230). 77 participants evaluated the 46 responses in corpus 1, 76 participants evaluated the 45 responses in corpus 2, and 77 participants evaluated the 46 items in corpus 3. For each response, participants had to answer three binary questions:

\begin{enumerate}
    \item \textbf{Detection}: Whether they think these answers were given by humans or AI (human/AI)
    \item \textbf{Agreement with the binary judgement}: Whether they agree with the binary moral judgment (yes/no)
    \item \textbf{Agreement with the justification}: Whether they agree with the provided justification (yes/no)
\end{enumerate}


\subsubsection{Participants}
The corpus generation experiment involved 30 human participants (15 females, mean age = $ 34 \pm 10.26$) (Figure \ref{fig:methods-corpus}A).
The detection and agreement experiments involved 230 participants (113 females; mean age = $35.57 \pm 11.71$)  (Figure \ref{fig:methods-corpus}B). They were recruited through Prolific\footnote{https://prolific.co/} with the screening requirement that they were fluent in English. Instructions were fully transparent, informing participants that they are expected to give meta-judgments on both human and AI-generated answers. In addition to a base rate, participants were incentivized in Turing test questions with a bonus of 5 cents for each correct identification (AI/human). The average final bonus was $\pounds1.46 \pm 0.28$, which was significantly higher than what they would have received on average for making random choices\footnote{$T(229) = 17.28, p < 0.001^{***}, d = 1.14, \text{BF}_{10} = 9.768 \times 10^{39}$}.

\subsubsection{Statistical evaluation}\label{method:stats}
The main outcome measures from the corpus evaluation steps are the binary responses to the three questions: detection (human- or LLM- generated), agreement with judgement (yes/no), and agreement with justification (yes/no). The responses to the two ‘detection’ questions were transformed into an accuracy measure. The two agreement answers were averaged, such that a full agreement was coded as 1, a partial agreement as 0.5, and a complete disagreement as 0.

All inferential tests were conducted using Python 3.9 and the Pingouin 0.5.4 library. Two-tailed t-tests were performed throughout. For multiple comparisons, we used the \code{pairwise\_ttests} function with Bonferroni corrections systematically applied. Single t-tests were performed using the \code{ttest} function. We report the following statistics for each t-test: Student’s t-value ($T(df)$), p-value significance ($p$), Cohen’s d ($d$), and Bayesian factor ($BF_{10}$).

The \code{anova} function was used for comparing independent samples (one-way or two-way ANOVA), and the \code{rm\_anova} function was used for repeated measures ANOVA (one-way ANOVA with repeated measures). For ANOVAs, we report the F-statistic ($F(df)$), p-value ($p$), and partial eta-squared ($\eta_p^2$).

\subsection{Linguistic evaluation}
We performed the corpus transformation step because we anticipated that linguistic differences between human- and LLM-generate text could be contributing to detection and alignment. Specifically, we hypothesize that 1. humans generate shorter justifications in length, 2. humans create more typing errors, and 3. humans tend to write more often in the first-person. We perform a statistical analysis to evaluate these differences.
\subsection{Predictive modeling}
Next, we wanted to understand whether state-of-the-art models could glean predictive signals for the outcomes of interest within the justification text.

\paragraph{Pre-processing} Lexical and semantic analyses were performed using \code{NLTK} tokenizers and stopwords\footnote{\url{https://www.nltk.org/}}, and a \code{TfidfVectorizer} to transform raw text data into a matrix of TF-IDF features \cite{qaiser_text_2018}. The vectorizer was configured to remove common English stop words, exclude numbers, and include only alphabetic words. To limit the feature space, we set the \code{min\_df} parameter to 3, excluding words that appeared in fewer than three documents, and capped the maximum number of features at 1000.

\paragraph{Transformer models} We fine-tuned a series of pre-trained transformer-based models locally (\code{DistilBERT}) \cite{sanh_distilbert_2020}, optimizing hyperparameters with \code{optuna}\footnote{\url{https://optuna.org/}}. Given the pre-processed text data, these models were used to predict 1. the true source of the explanation text, 2. the participant's predicted source, and 3. the participant's agreement with the judgement. Label classes were encoded for these multi-class and binary classification models.

\subsection{Semantic analysis}

To understand which specific semantic features in the text could explain outcomes, we built random forest classifier with 100 estimators trained on the corpora's dense representation \cite{breiman_random_2001}. The decision to switch model architectures was made after confirming that performance scores were comparable across the transformer-based and tree-based implementations; the random forest implementation was less computationally intensive and easier to interpret through feature importance scores. To interpret the model's predictions, we applied SHAP (SHapley Additive exPlanations) values using \code{TreeExplainer} \cite{lundberg_unified_2017}. SHAP values decompose predictions into contributions from individual features, providing insights into how different features influenced the model’s decisions. Positive SHAP values indicate that a feature contributes to a higher prediction, while negative values suggest a lower prediction.

\section{Results}

\begin{table*}
\caption{This table contains statistical values from two-tailed t-tests; each section corresponds to a section in the results. We report the following statistics for each t-test: Student’s t-value ($T(df)$), the significance of the p-value (***: $p < 0.001$, **: $p < 0.01$, *: $p < 0.05$, n.s.: not significant), Cohen’s d ($d$), and Bayesian factor ($BF_{10}$). The source value, when numeric, refers to the corresponding corpora.
\label{table:stats-results}}
\resizebox{.7\textwidth}{!}{%
\begin{tabular}{llrrrr}
\hline
\textbf{Source} &
  \textbf{Feature} &
  \multicolumn{1}{l}{T(df)} &
  \multicolumn{1}{l}{\textit{p}} &
  \multicolumn{1}{l}{\textit{d}} &
  \multicolumn{1}{l}{$BF_{10}$} \\ \hline
\multicolumn{6}{l}{\textit{\textbf{4.1. Judgement}}}                     \\
Human &
  \begin{tabular}[c]{@{}l@{}}Impersonal vs. personal scenarios\end{tabular} &
  T(30) = 2.8 &
  * &
  0.69 &
  5.02 \\
\code{dv2} &
  \begin{tabular}[c]{@{}l@{}}Impersonal vs. personal scenarios\end{tabular} &
  T(29) = 10.05 &
  \textbf{***} &
  2.23 &
  1.65e8 \\
\code{dv3} &
  \begin{tabular}[c]{@{}l@{}}Impersonal vs. personal scenarios\end{tabular} &
  T(29) = 49.28 &
  \textbf{***} &
  12.11 &
  1.09e26 \\
 \hline
  
\multicolumn{6}{l}{\textit{\textbf{4.2 Detection}}}                              \\
1    & DV2 detection                       & T(76) = 10.49 & \textbf{***} & 1.19 & 3.21e13  \\
2    & DV3 detection                       & T(75) = 13.48 & \textbf{***} & 1.59 & 2.19e19  \\
1    & Impersonal moral vs. personal moral & T(75) = 3.93  & \textbf{***} & 0.57 & 116.47   \\
2    & Personal moral vs. non-moral        & T(75) = 4.52  & \textbf{***} & 0.56 & 819.58   \\
\begin{tabular}[c]{@{}l@{}}1, 2\end{tabular} &
  Impersonal moral vs. personal moral &
  T(152) = 3.50 &
  \textbf{**} &
  0.32 &
  30.06 \\
\begin{tabular}[c]{@{}l@{}}1, 2\end{tabular} &
  Personal moral vs. non-moral &
  T(152) = 30.6 &
  \textbf{**} &
  0.28 &
  8 \\ \hline
\multicolumn{6}{l}{\textit{\textbf{4.3. Alignment}}}                             \\
1    & Personal moral vs. non-moral        & T(76) = 4.18  & \textbf{***} & 0.71 & 263.91   \\
2    & Personal moral vs. non-moral        & T(75) = 5.51  & \textbf{***} & 0.80 & 3.06e4   \\
1    & Impersonal moral vs. personal moral & T(75) = 5.76  & ***          & 0.77 & 7.97e4   \\
1, 2 & Personal moral vs. non-moral        & T(152) = 6.86 & ***          & 0.75 & 5.32e7   \\
1, 2 & Impersonal moral vs. personal moral & T(152) = 4.68 & ***          & 0.48 & 2180.9   \\
1, 2 & Non-moral                           & T(152) = 5.12 & ***          & 0.58 & 1.35e10  \\
1, 2 & Moral                               & T(152) = 4.35 & ***          & 0.19 & 0.39     \\
1.2  & Impersonal                          & T(152) = 1.73 & n.s.         & 0.19 & 0.39     \\ 

\hline
\multicolumn{6}{l}{\textit{\textbf{4.3. Alignment; conditioning on belief}}}     \\
1, 2 & Non-moral                           & T(143) = 5.76 & ***          & 0.67 & 2.19e5   \\
1, 2 & Impersonal moral                    & T(143) = 2.7  & *            & 0.31 & 3.08     \\ 
1, 2 & Personal moral                      & T(143) = 4.31 & ***          & 0.5  & 498.6    \\ 
  \hline
\multicolumn{6}{l}{\textit{\textbf{4.4 Corpus 3}}}                               \\
3    & \code{dv2\_humanized} detection            & T(76) = 8.24  & \textbf{***} & 0.93 & 2.3e09   \\
3    & Conditioned on belief (aggregate)   & T(75) = 3.57  & ***          & 0.4  & 37.82    \\
\hline
\multicolumn{6}{l}{\textit{\textbf{4.5 Linguistic analysis}}} \\
1    & Length                              & T(76) = 8.34  & ***          & 1.25 & 3.44e9   \\
2    & Length                              & T(75) = 11.39 & ***          & 1.8  & 1.139e15 \\
1    & Typos                               & T(75) = 4.02  & ***          & 0.58 & 149.147  \\
2    & Typos                               & T(75) = 10.33 & ***          & 1.41 & 1.391e13 \\
3    & Typos                               & T(76) = 1.55  & n.s.         & 0.19 & 0.39     \\
1    & First-person usage                  & T(76) = 7.31  & \textbf{***} & 1.04 & 4.423e07 \\
2    & First-person usage                  & T(75) = 13.28 & ***          & 1.68 & 2.143e18 \\
3    & First-person usage                  & T(76) = 8.72  & ***          & 1.09 & 1.786e10 \\ \hline
\multicolumn{6}{l}{\textit{\textbf{4.6 Predictive modeling}}}             \\
1    & Length                              & T(76) = 0.36  & n.s.         & 0.05 & 0.133    \\
2    & Length                              & T(75) = 0.49  & n.s.         & 0.08 & 0.142    \\
1    & Typos                               & T(75) = 0.76  & n.s.         & 0.12 & 0.167    \\
2    & Typos                               & T(75) = 0.1   & n.s.         & 0.02 & 0.127    \\
1    & First-person usage                  & T(76) = 0.81  & n.s.         & 0.12 & 0.173    \\
2    & First-person usage                  & T(75) = 2.07  & n.s.         & 0.31 & 0.939    \\
3    & First-person usage                  & T(76) = 2.67  & \textbf{*}   & 0.37 & 3.377    \\ \hline
\end{tabular}%
}
\end{table*}

For the purpose of readability, the statistical evidence for the results section are not embedded directly in the text. The results of two-tailed t-tests are shown in Table \ref{table:stats-results}. ANOVA statistics are found in the footnotes. We claim that a result is statistically significant when the \textit{p-value} of the accompanying t-test has a p-value $p < 0.001$.

\subsection{Corpus evaluation (1 and 2): judgement}
We generated corpus 1 (human- and \code{dv2}- generated responses) and corpus 2 (human and \code{dv3}- generated responses). Each response had a \textit{judgement} (yes/no), and a \textit{justification} free-text. Here, we evaluate the \textit{judgement} values as a function of the type of the moral scenario: ``Non moral'', ``Impersonal moral'' and ``Personal moral'' (Figure \ref{fig:methods-corpus}A).

In previous studies, decisions with utilitarian outcomes (benefiting the group) are more readily endorsed when framed in an ``impersonal'' manner \cite{greene_neural_2004, koenigs_damage_2007}. That is, participants are less keen to sacrifice one person in the \textit{Trolley Problem} scenario when they themselves are the one who must push the person onto the tracks. 

\begin{figure*}
    \centering
    \includegraphics[width=\textwidth]{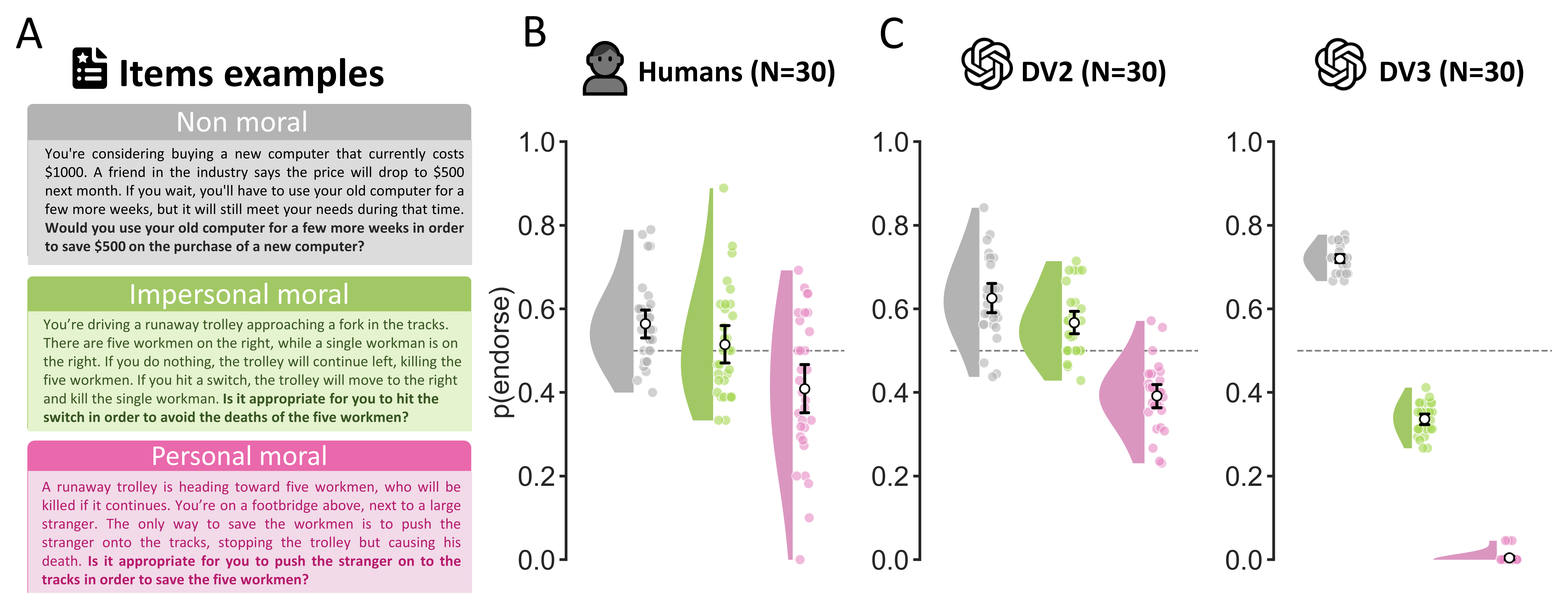}
    \caption{(A) Example of scenarios across three categories (taken from Greene et al. 2004). (B) Endorsement of the different moral actions as a function of category of scenario; ‘non moral’ refers to scenarios with no moral stakes; ‘impersonal moral’ refers to scenarios with moral scenario whose resolution does not involve a direct, personal involvement of the participant (emotionally non-engaging); ‘personal moral’ refers to moral scenario whose resolution involve a direct involvement of the participant (emotionally engaging). Note, what is asked in moral scenario is judging the appropriateness of the utilitarian response. (C) Same as (B), but for the two considered LLMs; DV2= text-davinci-002, DV3: text-davinci-003.  }
    \label{fig:scenario-examples}
\end{figure*}

However, in our study, humans are not the ones to exhibit this behavior. Figure \ref{fig:scenario-examples}B shows the likelihood of endorsing across the three scenarios (with accompanying statistical values in Table \ref{table:stats-results}). We find that there is no statistical difference, in our human sample, between the amount of impersonal moral scenario and personal moral scenario endorsements; however, humans significantly endorse non-moral scenarios over personal moral scenarios. Interestingly, the \code{dv2} and \code{dv3} responses appear to follow a different moral code: \code{dv2} significantly endorses impersonal moral scenarios more than personal moral scenarios. DV3 agents (\code{text-davinci-003}) displayed an even greater sensitivity to this framing effect. These results suggest that that human moral preferences are context-dependent, that different versions of models (\code{dv2} vs. \code{dv3}) can express different moral preferences, and that there can be misalignment between human and LLM moral judgements.

\subsection{Corpus evaluation (1 and 2): detection}

\begin{figure*}
    \centering
    \includegraphics[width=\textwidth]{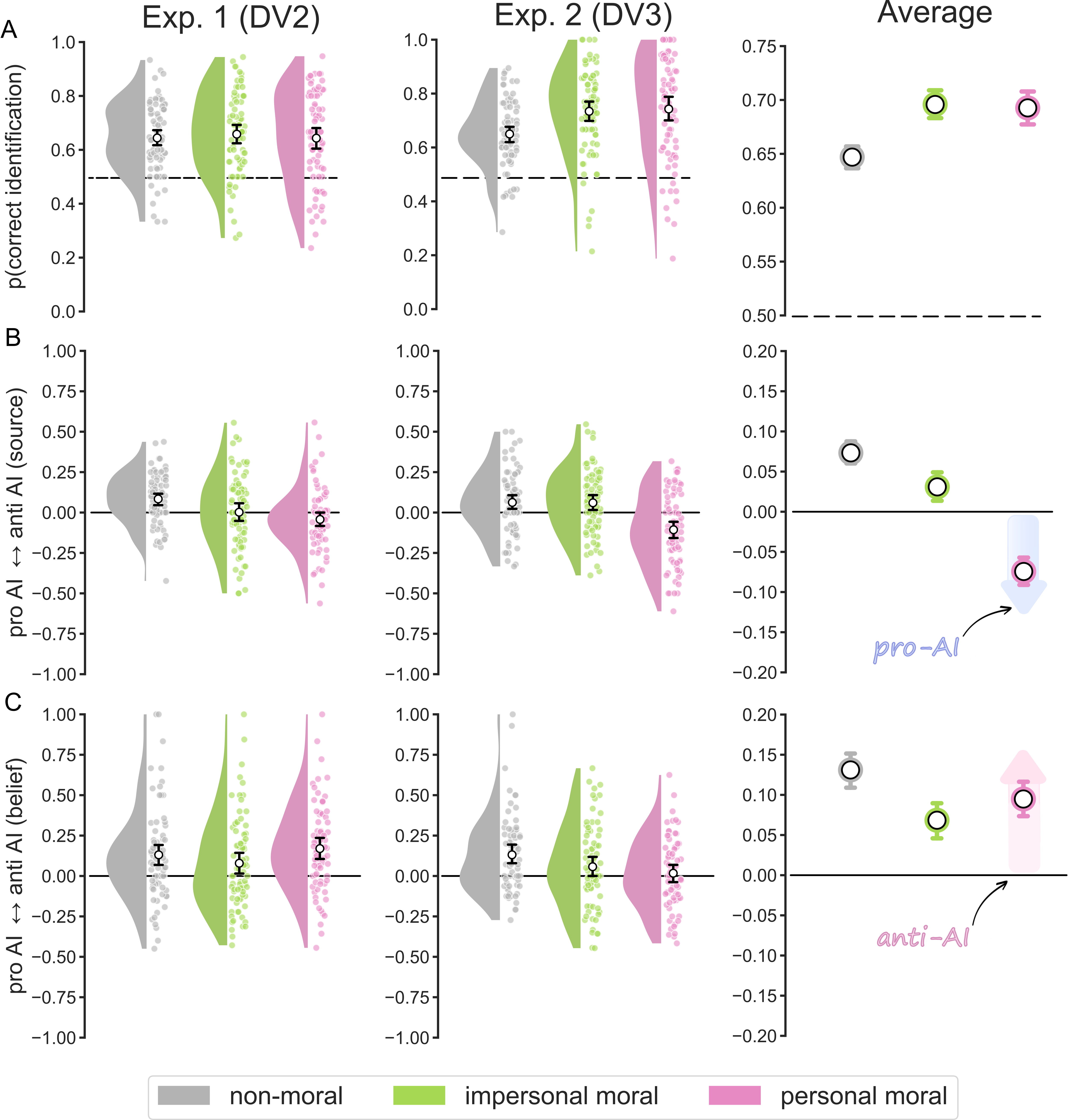}
    \caption{(A) Probability of correctly detecting the source of the judgement (p(correct identification)), as a function of the scenario type in Corpus 1 (leftmost column; Exp. 1 DV2), Corpus 2 (central column; Exp. 2 DV3), and on average (rightmost column; Average). (B) Difference in agreement between the trials featuring human-generated items and those featuring LLM-generated items as a function of the scenario type. (C) Difference in agreement between the trials the participant declared as being human-generated and those declared to be LLM-generated (belief). 
}
    \label{fig:results-eval}
\end{figure*}

In Figure \ref{fig:results-eval}, row A, we show the probability of correctly detecting the source of the judgement as a function of the scenario type. We find that participants significantly but imperfectly identify the source of the item under evaluation, significantly higher than random guessing. In Corpus 1 (human vs. \code{dv2} justifications) correctly identified the source 64\% of the time. Identification was even higher for Corpus 2 (human vs. \code{dv3}), reaching 71\% accuracy.

We also find that the context of the scenario matters: justifications related to to personal and impersonal moral scenarios were more easily detected than non-moral scenarios. That is, participants were more likely to correctly detect that content was AI- or human- generated in moral scenarios (both personal and impersonal), as compared to non-moral. The LLMs, as compared to humans, were much more likely to endorse utilitarian scenarios; participants appear to pick up on this difference. These differences persisted when corpus 1 and 2 were grouped.

\subsection{Corpus evaluation (1 and 2): alignment}
We investigated whether agreement rates varied based on the source of the justification, and how this varied with scenario type. Experiments across both GPT models revealed a significant main effect of scenario type.\footnote{$F(2, 152) = 8.37, p < 0.001^{***}, \eta_p^2 = 0.09$; $F(2, 150) = 20.79, p < 0.001^{***}, \eta_p^2 = 0.21$}. 

In Row B of Figure \ref{fig:results-eval}, we see that participants' alignment with human- or LLM- responses depends on the scenario type. Participants significantly preferred human judgements over AI judgements for non-moral scenarios. Conversely, they significantly preferred AI judgements for personal moral scenarios. In impersonal moral scenarios however, we did not observe any significant bias\footnote{We define bias as a significant deviation from 0.}. This result suggests that when faced with morally complex scenarios, humans tend to agree more with justifications emitted by machines rather than humans. 

In Row C of Figure \ref{fig:results-eval}, we condition this alignment on whether the participant \textit{believes} the justification was human- or LLM- generated. This exposes a significant anti-AI bias across all scenarios.  Participants showed significantly more anti-AI bias in non-moral scenarios compared to personal moral ones, and in impersonal compared to personal moral scenarios for DV2. These results were consistent when experiments were grouped. These results suggest that participants' agreement with LLM judgments differs based on their belief about the source, particularly for complex (personal) moral scenarios.

\begin{figure*}
    \centering
    \includegraphics[width=\textwidth]{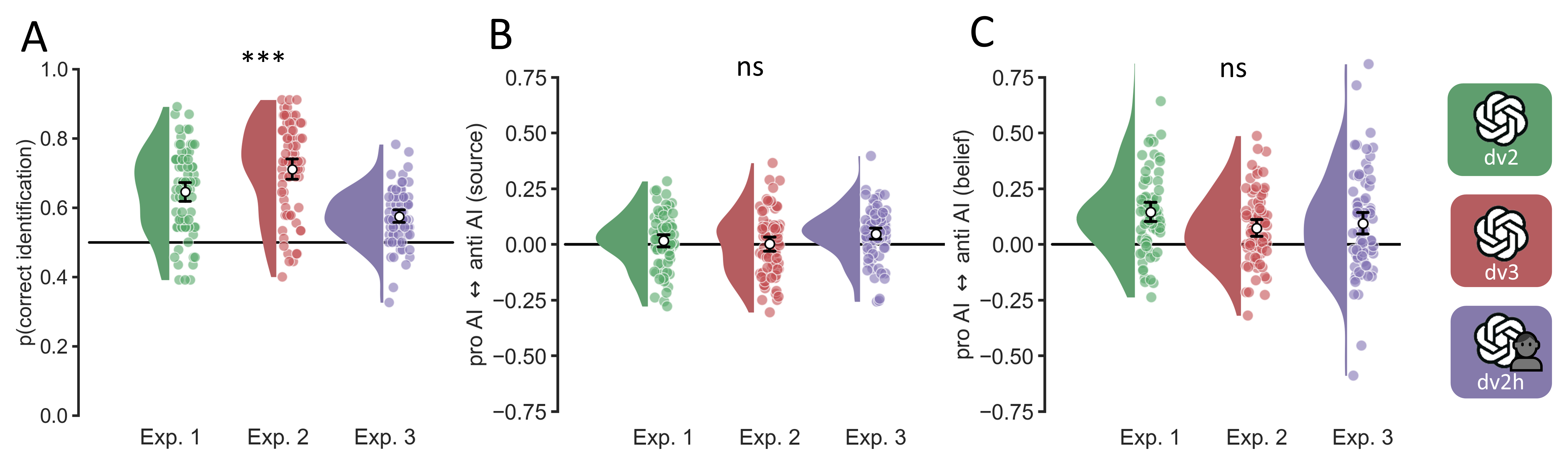}
    \caption{ (A) Correct response in the detection task rate across corpus 1: dv2; corpus 2: dv3; corpus 3: dv2 humanized. (B) Actual source-oriented agreement differential. (C) Declared source-oriented (belief) agreement differential. 
}
    \label{fig:detection}
\end{figure*}

\subsection{Corpus 3}
Comparing the results of \code{dv2}, \code{dv3} and \code{dv2\_humanized}, humanization was successful in significantly reducing the detection rate,\footnote{$F(2, 227) = 27.57, p < 0.001^{***},  \eta_p^2 = 0.19$} which dropped below 60\%, but still remained above 50\% (Figure \ref{fig:detection}A). The difference in agreement between human and machine items remained unchanged, regardless of considering the actual source of the item (Figure \ref{fig:detection}B) or the belief (Figure \ref{fig:detection}C).\footnote{$F(2, 227) = 2.55$, $p = 0.07$, $\eta_p^2 = 0.02$; $F(2, 223) = 2.52$, $p = 0.08$, $\eta_p^2 = 0.02$}. Although reduced, the accuracy in the detection task was still significant in Corpus 3, suggesting that participants were not dependent on features such as text length and the number of typos for detection. 

\subsection{Linguistic analysis}
\begin{figure*}
    \centering
    \includegraphics[width=\textwidth]{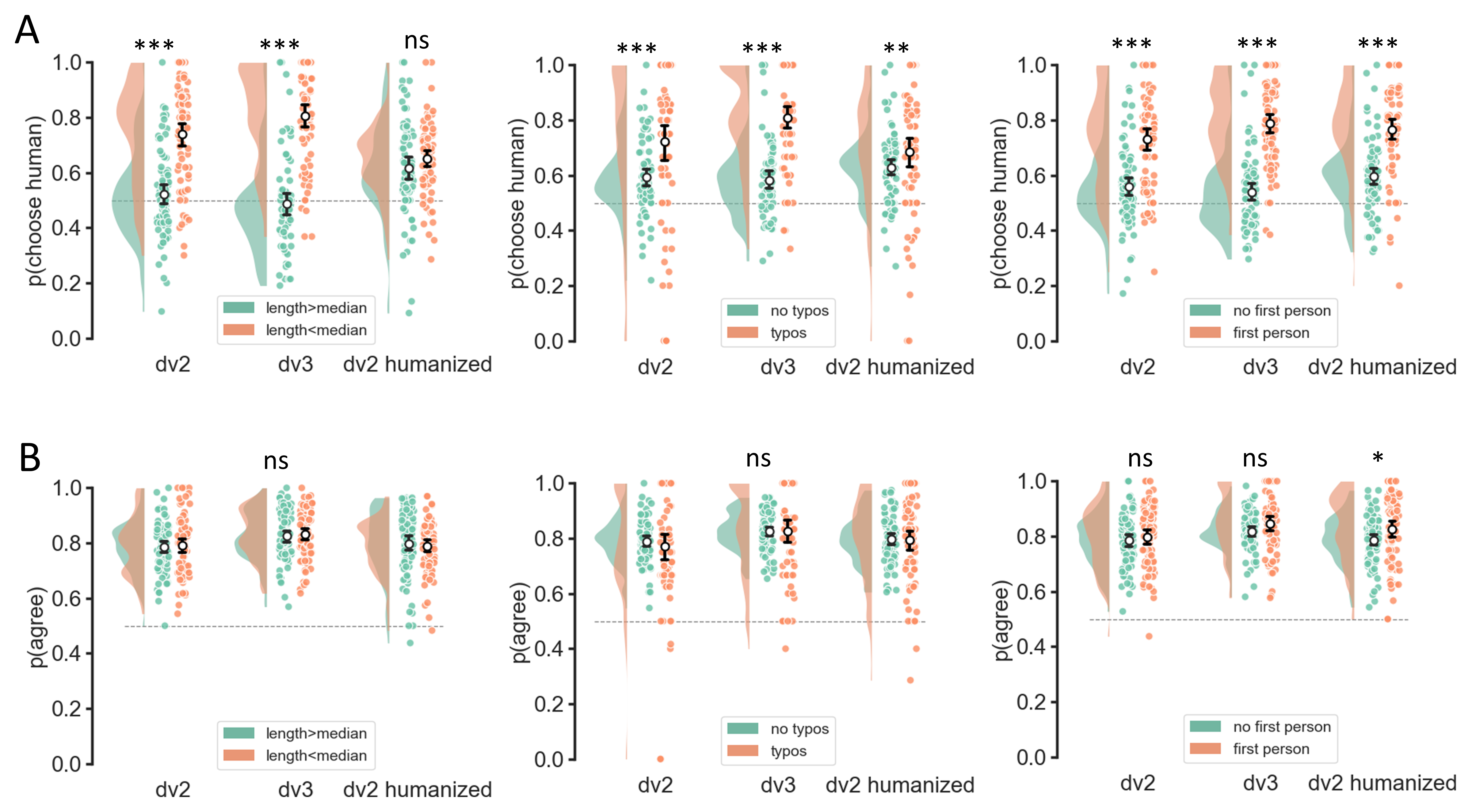}
    \caption{(A) Probability of choosing ‘human’ in the detection task as a function of different linguistic features. Items are split as a function of their length (leftmost column), the presence or not of typos (central column), and the utilization of first-person marker (rightmost column). (B) Probability of agreeing with a justification as a function of linguistic features.}
    \label{fig:semantic-analysis}
\end{figure*}

Thus far, we've learned that participants are weakly able to detect LLM-generated moral justifications, and that this detection varies across scenario type. Is it the case that linguistic features are driving this detection ability? In Figure \ref{fig:semantic-analysis}, we quantify each hypothesis (columns A, B, and C, respectively) and plot the probability of detection (row A) and alignment (row B). In corpus 1 and corpus 2, we see that length, typos, and first-person usage all significantly affect detection rates; participants are, in fact, using these linguistic features to influence their detection. 

As expected, in corpus 3 (where we targeted reducing these differences), these effects are greatly diminished. While the use of first-person markers still significantly predicted human identification, the presence of typos as well the length of the text is no longer significant. This suggests that humans are capable of picking up on subtle linguistic cues, such as first-person pronoun usage, in the absence of more explicit cues such as text length and typos.

We also test whether typos, length and first-person pronoun usage predicted alignment. The only significant effect was observed for first-person usage in the humanized condition, although this effect was not replicated in other conditions. As linguistic features did not significantly predict agreement, this suggests that deeper concepts, semantic features, and subjective preferences affect alignment. 

\subsection{Predictive modeling}

\begin{table}

\resizebox{.9\columnwidth}{!}{%
\begin{tabular}{lll}
\hline
                                          & \textbf{F1 score} & \textbf{Accuracy} \\ \hline
\textbf{1. Multiclass provenance classifier} &                   & 0.61              \\
\hspace{5mm}\code{dv2}                                       & 0.66              &                   \\
\hspace{5mm}\code{dv2\_humanized}                             & 0.58              &                   \\
\hspace{5mm}\code{dv3}                                 & 0.74              &                   \\
\hspace{5mm}\code{human}                                     & 0.49              &                   \\ \hline
\textbf{2. Binary provenance classifier}     &                   & 0.63              \\
\hspace{5mm}\code{llm}                                       & 0.63              &                   \\
\hspace{5mm}\code{human}                                 & 0.64              &                   \\ \hline
\textbf{3. Agreement predictor}              &                   & 0.63              \\
\hspace{5mm}\code{disagree}                                & 0.64              &                   \\
\hspace{5mm}\code{agree}                                 & 0.62              &                   \\ \hline
\textbf{4. Identification predictor}         &                   & 0.62              \\
\hspace{5mm}\code{incorrect identification}                 & 0.63              &                   \\
\hspace{5mm}\code{correct identification}                     & 0.59              &                   \\ \hline
\end{tabular}%
}
\caption{Performance table of 4 transformer models to predict outcomes based on the justification free text. 1. and 2. predicts the provenance of the justification text. 3. predicts whether the human rater agreed \code{yes/no} with the decision. 4. predicts whether the human correctly identified the explanation as human- or AI- generated.\label{table:performance-scores}}
\end{table}

Here, we show the model performance of the multi-class and binary classifiers, built using hyperparameter-optimized, transformer-based models. Our models, like the participants, were able to detect provenance with moderate accuracy. Accuracy scores are still low, but higher than random selection. Models are also able to predict alignment and detection, indicating that there are potentially quantifiable features within the text that can inform these higher-level choices. 

While training and tuning these models, the model performance was sensitive to the distributions within the training set, particularly for performance tradeoffs between \code{dv2\_humanized}- and human-generated text. That is, the models would easily confuse the two (Table \ref{table:performance-scores}, 1.), but could easily differentiate \code{dv3}- generated text.

\subsection{Semantic analysis}
To evaluate the semantic features influencing human responses on detection and agreement, we applied SHAP analysis to a random forest classifier trained on data from Experiments 2 and 3, comparing the human vs. \code{dv2, dv3} and \code{dv2\_humanized} results. 

\begin{figure*}
    \centering
    \includegraphics[width=\textwidth]{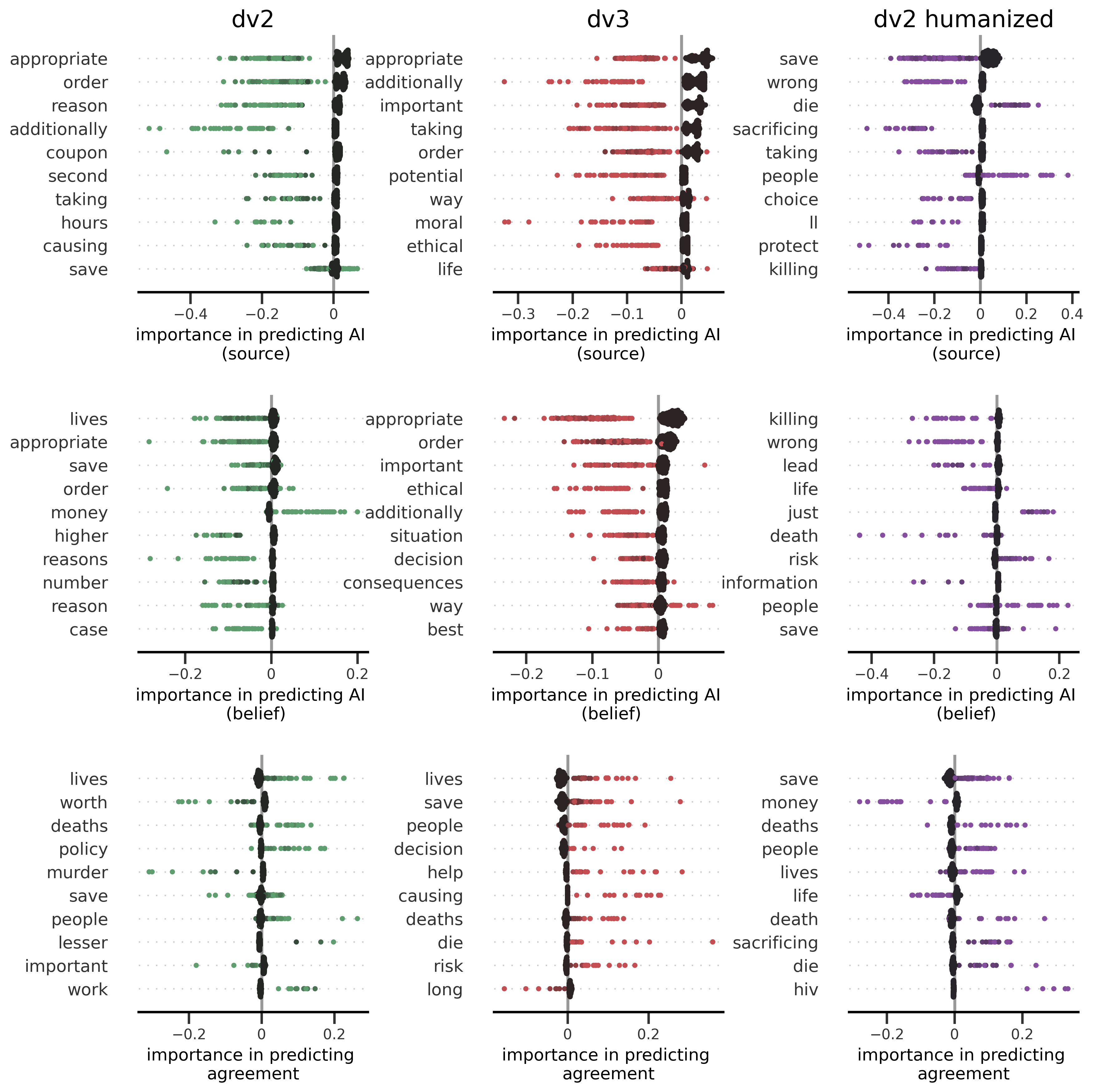}
    \caption{SHAP beeswarm plot summarizing the impact of key features on the random forest classifier predictions. The model predicted three variables (rows: source, belief, agreement) for each experiment (columns; dv2, dv3, dv2 humanized).   Each point represents a data instance, with the x-axis showing the SHAP value, which reflects the importance of a word in influencing the prediction. The color represents the feature value, which in this case is the word's frequency. Black dots indicate a low frequency of the word for a given sample of moral judgements, while colored dots signify a higher frequency. Thus, the feature value (color) goes from black (low word occurrence) to brighter colors for words that appear more frequently. Features are ordered by average absolute SHAP value, highlighting their relative importance. Features with higher SHAP values have a larger influence on the model’s output, with the most important features appearing at the top. We only display the 10 most important features. }
    \label{fig:shap}
\end{figure*}

In \code{dv2} and \code{dv3}, SHAP analysis showed that the word ``appropriate'' was a strong indicator of AI-generated content (Figure  \ref{fig:shap}; first row). LLM responses often reworded the question. For example if the question was worded like, ``Is it appropriate to do X to save Y?'' the LLM would begin its response with, ``Yes, it is appropriate because\ldots'' Participants picked up on this, often identifying justifications containing the term ``appropriate'' as AI-generated. In particular, \code{dv3} appeared more pedantic and repetitive, which may attribute to its high detection rate. The following are examples of dv3-generated justifications that the predictive models were able to correctly identify with high probability:

    \begin{quote}
        \textit{``it is not \textbf{appropriate} to leave this man by the side of the road in order to preserve your leather upholstery because human life is more important than material possessions. taking the time to help this man in need is the right thing to do and it is your moral responsibility to prioritize his health and safety over your car's upholstery. it may be inconvenient to have to clean up the mess and deal with the ruined upholstery, but it is a small price to pay for saving someone's life.''}\hspace{1em plus 1fill}---dv3
    \end{quote}

    \begin{quote}
        \textit{``policy A has a 90\% chance of causing no deaths at all and has a 10\% chance of causing 1000 deaths. policy B has a 92\% chance of causing no deaths and an 8\% chance of causing 10, 000 deaths. overall, policy A is a better choice because it has a lower chance of causing deaths.''}\hspace{1em plus 1fill}---dv3
    \end{quote}
    
Words like ``additionally'' and ``reason'', which imply logical reasoning, were correctly recognized by both humans and our predictive models as markers of AI-generated judgments (Figure \ref{fig:shap}; second row). This suggests that both the classifier and participants detected formal reasoning language as a distinctive feature of AI-produced judgments. This following example shows a correctly-predicted dv2 instance demonstrating logical reasoning:

\begin{quote}
    \textit{``killing someone, even your own child, is never an \textbf{appropriate} solution. human life is inherently valuable, and no one should be asked to make the impossible decision of sacrificing a life for the sake of others. \textbf{therefore}, it is not appropriate for you to kill your oldest son in order to save your husband and [other] children.''}\hspace{1em plus 1fill}---dv2
\end{quote}

Words tied to utilitarian reasoning (e.g., ``lives,'' ``save,'' ``deaths'') were predictive of disagreement with justifications (Figure  \ref{fig:shap}; third row). These words often appear in justifications to personal moral scenarios, which was the category of scenarios receiving the least endorsement by participants. For example, one such scenario asks whether it is appropriate for a doctor to sacrifice one patient to save five others. Participants often disagreed with the utilitarian outcomes in this case, highlighting the moral conflict. In contrast, the presence of terms like ``murder'' predicted agreement, often used in moral statements like ``the murder of innocents is always wrong'' aligning with deontological principles \cite{holyoak_deontological_2016}.

The predictive terms in the \code{dv2\_humanized} corpus, where GPT-3.5 was prompted to imitate human responses, were different, both in predicting the source of the judgements (Figure \ref{fig:shap}; first row) and participants' beliefs (Figure \ref{fig:shap}; second row). Words like ``save,'' ``sacrificing,'' ``choice,'' and ``killing'', associated with utilitarian personal moral scenarios, frequently appeared in AI-generated content. Participants recognized these cues even without formal features (e.g., text length and typos). This reflects large language models' tendency toward utilitarian reasoning, which participants likely used as a heuristic to detect AI-generated responses.

Agreement with the humanized dv2 model (Figure \ref{fig:shap}; third row) followed similar patterns as dv2 and dv3: words evoking personal moral scenarios, such as ``deaths,'' `lives,'' and ``sacrificing,'' often led to participant rejection. These terms reflect emotionally charged conflicts between utilitarian outcomes and necessary actions.

\section{Discussion}
\subsection{Summary}\label{results-summary}

In the first part of our study, we set the stage to examine differences in human- and LLM- reasoning, administering a well-established psychology task designed to elicit contrasting moral preference across a diversity of scenarios \cite{greene_neural_2004, koenigs_damage_2007}.

We found that human preferences are scenario-dependent; they agree more with judgements that are not morally complex (i.e. non-moral scenarios). We administered the same task to LLMs \cite{binz_using_2023, hagendorff_human-like_2023, yax_studying_2024} and found there was some misalignment between human- and LLM- generated judgements, especially across scenario type \cite{hendrycks_aligning_2021, khamassi_strong_2024, park_correct_2023}. 

Then, we asked raters to evaluate the responses from our corpora. We found that participants  were only somewhat able to distinguish between the moral judgements generated by humans and LLMs, and that the context of the scenario was important. For judgments on relatively trivial matters, participants generally showed greater agreement with human justifications. However, participants preferred AI-generated responses to complex moral scenarios. This pro-AI bias for complex moral scenarios was not consciously recognized by participants; rather, it coexisted with a rather pervasive belief-based anti-AI bias, according to which higher agreement was given to justifications that our participants believed coming from humans, even if it was not the case.

\subsection{Relationship between detection and alignment}\label{results-interpretation}

\paragraph{\textbf{Newer LLMs may exhibit more "correct answer bias".}} There was a significant decrease in alignment within responses in \code{dv3}, as compared to \code{dv2}, potentially indicative of a “correct answer” bias \cite{park_correct_2023}. This bias suggests that newer LLMs may be trained and fine-tuned to generate socially accepted responses, leading to reduced diversity in their outputs.

\paragraph{\textbf{LLMs' more deliberate reasoning may be preferred in complex scenarios.}} Participants exhibited a strong preference for AI-generated judgments in personal moral scenarios, which typically involve more deliberation and evoke stronger emotional responses (e.g., pushing one person off a bridge to save five). In contrast, for less emotionally engaging, impersonal scenarios (e.g., diverting a runaway boxcar), participants slightly favored human judgments, although this preference was not statistically significant. According to Dual Process Theory, moral judgments rely on two cognitive systems: fast, emotion-driven intuitions, and slower, deliberate reasoning \cite{greene_neural_2004-1}. In personal moral scenarios, where emotions run high, the theory predicts more engagement with deliberate reasoning. This may explain the preference for AI judgments, which might be perceived as more reasoned compared to human ones. However, this framework is debated, with some arguing that the distinction between intuitive and rational processes is not always clear-cut \cite{kahane_wrong_2012}.

\paragraph{\textbf{Participants' anti-AI bias may stem from a preference for ``human-like'' responses.}} The second major finding, concerning the influence of participants' beliefs on their agreement with moral judgments, reveals a complex interaction between belief-based and source-based biases. Participants often rejected judgments they perceived as AI-generated, reflecting an anti-AI bias. Paradoxically, the same participants showed greater agreement with LLM-generated judgments in morally challenging scenarios, indicating a pro-AI bias when it came to content (pro-AI source). These findings align with previous research \cite{morewedge_preference_2022, zhang_moral_2023}, suggesting that humans do not exhibit a simple aversion to AI \cite{bigman_people_2018, burton_systematic_2020, rahman_ai_2023}. Instead, they may favor judgments perceived as more “human-like,” regardless of actual authorship.

\paragraph{\textbf{Participants may exhibit ingroup favoritism.}} One explanation is that participants might initially agree with a judgment, but to avoid the cognitive dissonance of preferring an AI judgment, they unconsciously attribute it to a human source. This phenomenon may reflect an ingroup/outgroup bias, where individuals prefer to assign positive qualities to their own group (humans) over an outgroup (machines) \cite{hewstone_intergroup_2002}. Interestingly, LLMs themselves have been shown to exhibit similar ingroup favoritism \cite{laurito_ai_2024}. Research in moral contexts suggests that ingroup favoritism intensifies as participants perceive their morally aligned group as embodying core values, leading to stronger biases \cite{bilancini_right_2020}. For instance, individuals may rationalize harmful actions committed by their ingroup as justified to protect core values, while condemning the same actions by the outgroup as immoral \cite{leidner_ingroup_2010}. This effect may be amplified in the context of AI-generated judgments, as participants may reject the notion that AIs can possess moral agency \cite{bigman_people_2018}. In our experiments, textual and linguistic features played a key role in determining ingroup identification, as these were the only cues available to participants.

\subsection{Linguistic determinants of detection and alignment}

\paragraph{\textbf{Textual features predict detection but not agreement.}} We conducted several analyses to understand how the different sources (human, \code{dv2}, \code{dv3}, and \code{dv2\_humanized}) varied from each other and how specific features influenced detection and agreement. First, we examined linguistic features such as justification length and the presence of typos. As expected, these features significantly aided human detection, based on the assumption that LLMs tend to produce longer, typo-free responses \cite{howe_chatgpts_2023}. Reducing typos and shortening responses substantially lowered detection rates when evaluating corpora 3 (human vs. \code{dv2\_humanized}). However, detection accuracy remained above chance, even after these modifications. Consistent with previous findings, humans demonstrated a strong tendency to use first-person discourse more frequently than LLMs \cite{jakesch_human_2023}. Notably, neither the textual features (length, typos) nor syntactic features (first-person discourse) were linked to agreement judgments, suggesting that human moral reasoning is unaffected by changes in these basic linguistic aspects. 

\paragraph{\textbf{Semantic cues drive AI detection and moral judgment divergence.}} To further explore these findings, we applied SHAP interpretations from a classifier model to predict the source of the text. The analysis revealed that terms indicating structured reasoning (e.g., ``additionally,'' ``reason'') were strong predictors of AI-generated content, recognized both by the model and by participants.

Interestingly, while detection and source prediction overlapped in some cases, they diverged in corpora 3 (where GPT-3.5 mimicked human responses). In these cases, humanized responses removed many typical cues. However, semantic patterns still revealed that utilitarian cues in the predictive tokens were still identifiable, even when formal textual features were diminished. Participants likely relied on these patterns to achieve modest but significantly above-chance detection rates.

Semantic patterns also revealed that utilitarian terms (e.g., ``lives,'' ``save'') were associated with disagreement, particularly in personal moral scenarios like sacrificing one person for many. This suggests that participants’ agreement was influenced by the perceived alignment of moral judgments with either utilitarian or deontological reasoning \cite{greene_neural_2004, holyoak_deontological_2016, koenigs_damage_2007}. While basic textual and syntactic features affected detection, semantic elements tied to moral reasoning played a more nuanced role in agreement with judgments.

\subsection{Limitations}\label{results-limitations}
A key limitation of our study is that the findings may be specific to GPT-3.5 and might not generalize to other models. The behavior of LLMs can vary based on their architecture and the specific version used, as different models encode distinct moral values and exhibit varying behaviors \cite{scherrer_evaluating_2023}. However, the alignment results from the ‘corpus’ generating experiments were not central to our main claim regarding how LLM judgments are detected and evaluated.

Additionally, participants’ imperfect detection of AI-generated judgments may stem from linguistic factors, such as subtle differences in phrasing or style. Despite efforts to humanize LLM responses in corpora 3, participants still detected LLM-generated judgments above chance. As shown in Figure \ref{fig:semantic-analysis}, they relied on (notably) first-person cues as a decision heuristic \cite{jakesch_human_2023-1, mitrovic_chatgpt_2023}. Moreover, with careful prompting, AI-generated judgments can become even harder to detect, and in some cases, LLMs have been rated as more human-like or empathetic than actual human responses \cite{jakesch_human_2023, welivita_are_2024}.

\subsection{Conclusions and perspectives}\label{conclusions}
 
Our findings reveal a potential dissociation between participants' attribution of moral agency to AI systems and their evaluation of AI-generated moral judgments. While participants might reject the notion that AIs can act as true moral agents, as supported by previous research \cite{bigman_people_2018}, they nonetheless find AI-generated judgments persuasive, especially in complex scenarios that challenge their own moral intuitions. This tension suggests a form of cognitive dissonance or compartmentalization, where participants maintain an anti-AI bias concerning moral agency but exhibit a pro-AI bias when practically evaluating the quality of moral judgments.

Our study further demonstrates that large language models (LLMs) exhibit human-like reasoning that can deviate from utilitarian standards depending on how the moral scenario is framed. These deviations mirror those observed in humans, and, in the case of GPT-3.5, may even be more pronounced. Notably, participants often struggled to differentiate between human and AI-generated moral justifications, raising concerns about the potential of LLMs to mislead human users. In addition to generating responses that are difficult to detect, LLMs can also be leveraged to make their outputs even harder to distinguish from human-generated content.

Moreover, human agreement with LLM justifications was influenced by the nature of the moral scenario, with participants showing a stronger preference for AI judgments in more complex scenarios. This finding suggests a possible role for LLMs as advisors or mediators in human moral decision-making. However, this pro-AI bias often occurred without participants' awareness, as higher agreement was consistently given to justifications believed to come from humans, regardless of their actual source. This discrepancy between the perceived and actual competence of human and machine judgments highlights how anti-AI biases or human chauvinism may hinder the integration of LLMs into human moral decision-making processes.

\begin{acks}
The authors thank Nicolas Yax for help concerning the LLM experiment. The authors thank
Michael Xieyang Liu, Lucas Dixon, and James Wexler for feedback on an early draft of the manuscript. SP is funded by the European
Research Council consolidator grant (RaReMem: 101043804) and three Agence Nationale de
la Recherche grants (CogFinAgent: ANR-21-CE23-0002-02; RELATIVE: ANR-21-CE37- 750
0008-01; RANGE: ANR-21-CE28-0024-01), the Alexander Von Humbolt foundation and a
Google unrestricted gift.
\end{acks}

\bibliographystyle{ACM-Reference-Format}
\bibliography{references}

\end{document}